\documentclass[12pt,twocolumn]{prop}

\pdfoutput=1

\usepackage[letterpaper,margin=1in]{geometry}

\usepackage{amsmath}

\usepackage[lf,mathtabular,footnotefigures]{MinionPro}
\makeatletter
 \def\Mn@Text@Family{MinionPro-TLF}
 
\makeatother

\usepackage[expansion=all,protrusion=default]{microtype}

\usepackage{graphicx}

\usepackage{cite}

\begin{document}

\begin{titlepage}
\mbox{}

\vfill

\begin{center}
  \Huge\bfseries
  Cosmological Studies \\ With A Large-Area X-ray Telescope
\end{center}

\vfill

\begin{center}
  \Large
  A.\,Vikhlinin\footnotemark[1],
  S.\,W.\,Allen\footnotemark[2],

  M.\,Arnaud\footnotemark[3],
  M.\,Bautz\footnotemark[4],
  H.\,B\"ohringer\footnotemark[5],
  M.\,Bonamente\footnotemark[6],
  J.\,Burns\footnotemark[7],
  A.\,Evrard\footnotemark[8],
  J.\,P.\,Henry\footnotemark[9],
  C.\,Jones\footnotemark[1],
  B.\,R.\,McNamara\footnotemark[10],
  D.\,Nagai\footnotemark[11],
  D.\,Rapetti\footnotemark[2],
  T.\,Reiprich\footnotemark[12]
\end{center}

\vfill

\centerline{\includegraphics[width=1.2\linewidth]{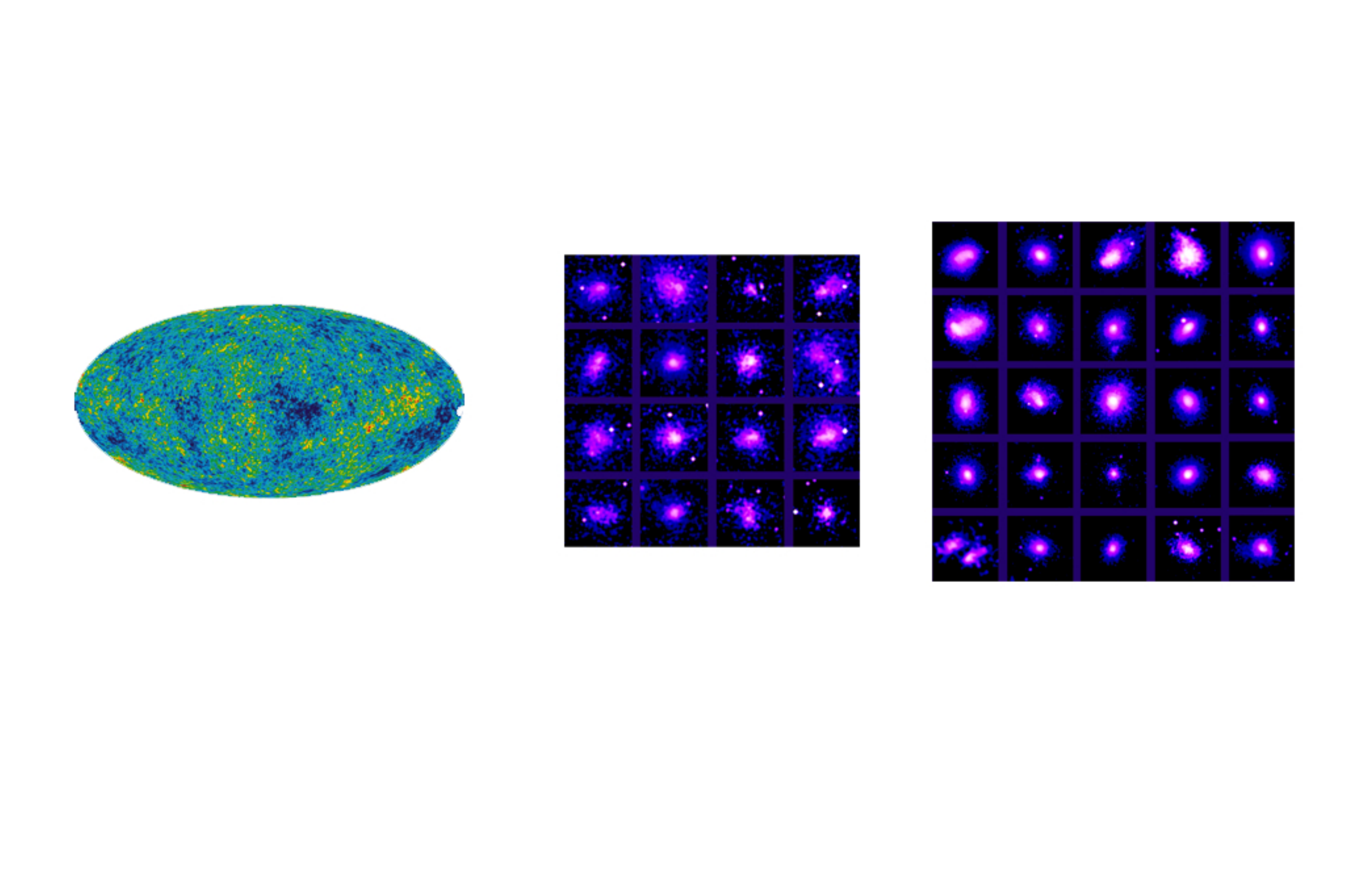}}

\vfill
\vfill

\mbox{}

\footnotetext[1]{Harvard-Smithsonian Center for Astrophysics}
\footnotetext[2]{Stanford University}
\footnotetext[3]{CEA/DSM/DAPNIA/SAP, CEA-Saclay}
\footnotetext[4]{MIT}
\footnotetext[5]{MPE}
\footnotetext[6]{University of Alabama}
\footnotetext[7]{University of Colorado}
\footnotetext[8]{University of Michigan}
\footnotetext[9]{University of Hawaii}
\footnotetext[10]{University of Waterloo}
\footnotetext[11]{Yale University}
\footnotetext[12]{University of Bonn}

\end{titlepage}

\section{X-ray cluster cosmology}

Clusters of galaxies are a very promising cosmological tools, in
particular because X-ray, SZ and optical and near-infrared data can be
combined to minimize systematic errors in identifying and characterizing
the cluster population. Mass selection systematics are smallest for
X-ray or SZ selection, and there is concrete hope for further reduction
in these uncertainties in the near future. X-ray astronomy in particular
has played an important role in establishing the current cosmological
paradigm. In the early 1990s, X-ray measurements of the baryonic mass
fraction in nearby galaxy clusters, coupled with improved measurements
of the Universal mean baryon density, provided some of the first
compelling evidence that we live in a low density Universe
\cite{1993Natur.366..429W}. Starting with early 1990's, X-ray
measurements of the local number density of clusters and its evolution
have also consistently pointed out towards a low-density Universe and a
relatively low value of amplitude of matter fluctuations, $\sigma_8$,
\cite{1991ApJ...372..410H,2002ApJ...567..716R,1998MNRAS.298.1145E,2001ApJ...561...13B,2003MNRAS.342..287A,2003A&A...398..867S},
a result since confirmed by cosmic microwave background (CMB) studies,
cosmic shear, and other experiments
\cite{2007ApJS..170..377S,2008arXiv0803.0547K,2008arXiv0811.4280D,2007MNRAS.381..702B,2008A&A...479....9F}.

Robust and precise understanding of dark matter and dark energy and how
they shape the structure and evolution of our Universe can be obtained
only through multiple, independent tests. The next generation of X-ray
observatories will, among many other things, provide powerful, new tools
to probe the structure and mass-energy content of the Universe. These
tools will be highly complementary to the best other planned
cosmological experiments (Planck, JDEM, LSST). In particular, the unique
capabilities of the International X-ray Observatory (IXO) will allow the
fullest possible exploitation of forthcoming cluster surveys made at
X-ray and other wavelengths, and enable the tightest possible control of
systematic uncertainties. Together, a powerful X-ray observatory and
these other experiments should enable a quantum leap in our
understanding of the Universe.

Cosmological studies in X-rays use observations of galaxy clusters.
X-ray data for clusters are crucial since $\sim 85\%$ of the baryons
within them are in the form of hot X-ray emitting gas. Precise
measurements of the X-ray brightness and temperature of this gas permit
two powerful and independent types of cosmological tests.

Firstly, observations with a powerful X-ray observatory such as IXO will
constrain the growth of cosmic structure, primarily by providing
accurate measurements for high-redshift galaxy clusters detected in new,
large X-ray and SZ surveys. The eROSITA or proposed WFXT X-ray missions,
for example, will discover $\sim\text{a few}\times10^5$ clusters within
$z\lesssim 2$, but provide only limited information on the individual
properties of high-$z$ objects.  Utilizing its much greater collecting
area and improved spatial and spectral resolution, IXO will provide
precise X-ray mass proxies for a complete subset of these clusters,
enabling a much tighter coupling between the survey fluxes and
theoretically predicted mass function
\cite{2001MNRAS.321..372J,2008ApJ...688..709T}. This will dramatically
enhance the power of these surveys to constrain cosmological parameters
\cite{2006ApJ...650..128K,2008arXiv0812.2720V}.  A large catalog of
serendipitously discovered clusters will further extend our knowledge of
clusters to fainter X-ray fluxes and higher redshifts, beyond $z=2$.

The second type of cosmological test possible at X-rays is primarily
geometric and, like type Ia supernovae (SNIa), constrains the expansion
history of the Universe, measuring distance-redshift relation, $d(z)$.
Here, the constraints will primarily come from measurements of the X-ray
emitting gas mass fraction, $f_{\rm gas}$, in the largest, most
dynamically relaxed galaxy clusters: $f_{\rm gas}$ is a
theoretically-predicted and observationally-verified `standard quantity'
associated with large clusters (see \cite{2008MNRAS.383..879A} and
references therein).  Additional, independent constraining power will
also be obtained from the combination of X-ray observations with
measurements of the SZ effect in the same clusters.

The ability of IXO to measure the primary X-ray observables (X-ray
brightness, temperature, metallicity and, for the first time, velocity
structure in high-$z$ objects) to exquisite precision in a large subset
of high-redshift clusters will, when coupled with external information
from state-of-the-art hydrodynamical simulations, gravitational lensing
studies and follow-up SZ observations, enable the tightest possible
control of systematic uncertainties in all cosmological measurements
using galaxy clusters.

\section{Measuring the growth of cosmic structure}

\noindent The main techniques proposed to study growth of cosmic
structure in future cosmological experiments are 1) measuring the
evolution of the mass function of galaxy clusters; 2) wide-area cosmic
shear surveys; and 3) using redshift-space distortions in the
galaxy-galaxy correlation function. The cosmic shear method is currently
the only growth of structure component of the proposed JDEM mission. At
present, the constraints on dark energy from cosmic shear and
redshift-space distortions are weak. In contrast, the constraints from
X-ray studies of the cluster mass function are relatively strong and
developing rapidly. The dominant systematic effects in the X-ray method
are clear and ways to address them have been identified and are being
vigorously pursued using e.g. follow-up gravitational lensing studies,
SZ observations, and hydrodynamical simulations.

Recent X-ray studies of the evolution of the cluster mass function using
the \emph{Chandra} X-ray Observatory to follow up ROSAT X-ray selected
clusters, have convincingly demonstrated that the growth of cosmic
structure has slowed down at $z<1$, due to the effects of dark energy.
These measurements have been used to improve the determination of the
equation state parameter \cite{2008arXiv0812.2720V} and to place first
constraints on possible departures from General Relativity
\cite{2008arXiv0812.2259R}. With IXO, working in concert with other
mutli-wavelength facilities to exploit new cluster surveys, it will be
possible to make similar measurements out to redshifts $z\sim2$,
providing a unique and critical insight into cosmic structure growth.
 
\subsection{The basics of cluster mass function measurements}

\noindent The mass function of galaxy clusters, $n(M)$, is an
exponentially sensitive indicator of the linear density perturbation
amplitude at the $\sim 10\,h^{-1}$~Mpc scale. Given precise cluster
masses, the perturbation growth factor in a given redshift bin can be
recovered to $1\%$ accuracy from a sample of only 100 clusters in the
$10^{14}-10^{15}$ solar mass range (for fixed values of all other
cosmological parameters). This high sensitivity is the main reason why
`counting clusters' provides such an attractive technique for studying
the growth of structure.

At present, cluster surveys provide a degenerate combination of
constraints on the growth of mass perturbations and the overall
geometric properties of the Universe. [This is simply because the volume
elements and masses derived from observations are both a function of
$d(z)$]. However, looking ahead a decade from now, we can expect
techniques such as SNIa, BAO and the X-ray $f_{\rm gas}(z)$ method to
have measured $d(z)$ with sufficient precision that the evolution of the
cluster mass function will become an essentially `pure' growth of
structure test. At this point, precise mass function measurements will
bring unique degeneracy-breaking power, powerfully and straightforwardly
enabling significant improvements in constraining the evolution of the
dark energy equation of state and in helping to distinguish the origin
of cosmic acceleration.

One of the most interesting applications for growth of structure data is
in testing theories that attempt to explain cosmic acceleration by
modifying the standard rules of gravity (General Relativity) on large
scales. In general, modifications to GR will affect theoretical
predictions for the cluster mass function by changing both the growth
rate of linear perturbations and modifying the process of non-linear
collapse. The process of non-linear collapse is already well calibrated
for GR-based dark energy models \cite{2008ApJ...688..709T}. As the field
of research matures, we can expect that the calibration will also become
similarly robust for other interesting non-GR models.  The combination
of precise X-ray and lensing, as well as SZ and optical-dynamical,
measurements of the baryonic and dark matter distributions in individual
clusters will also aid in probing modified gravity theories on
Megaparsec scales because in non-GR theories, dynamic and weak lensing
mass estimated in general are not expected to yield the same value.

\subsection{Strategy for mass function measurements}

\noindent Systematic, not statistical, uncertainties provide the
limiting factor in cosmological measurements based on the cluster mass
function. The cluster catalog provided by future X-ray survey missions
will be very large and, due to the strengths of X-ray techniques, should
have excellent purity and completeness. The primary need will be the
accurate calibration of cluster masses. Mass uncertainties are generally
of two kinds: 1) systematic $average$ biases in the derived masses; and
2) $scatter$ in the mass measurements for individual clusters. Both
sources of uncertainty are damaging.  For example, systematic
uncertainties of $\pm 10\%$ in the mean cluster mass at a given redshift
automatically lead to $\pm3\%$ uncertainties in the growth factor.

No single cluster mass measurement technique can address both
uncertainties. However, the $combination$ of X-ray and lensing methods
provides an approach that is both bias-free and has minimal intrinsic
scatter, and is also insensitive to detailed physics of cluster
formation.

X-ray hydrostatic analyses can provide low-scatter, and even relatively
low-bias, mass estimates for the most dynamically relaxed clusters.
However, for most systems, systematic scatter and biases in hydrostatic
mass estimates are expected at the $20-30\%$ level. Although IXO will be
able to measure and/or eliminate some such sources of uncertainty (e.g.,
by measuring bulk motions and turbulence in the intracluster medium via
high-resolution X-ray spectroscopy), controlling them at the few percent
level from X-ray data alone would appear to be impossible. However, {\it
  one does not require hydrostatic X-ray mass estimates for cluster mass
  function work}.  High-resolution cosmological simulations have shown
that the parameter $Y_X=T\times M_{\rm gas}$, where $T$ is the average
temperature derived from the X-ray spectrum and $M_{\rm gas}$ the gas
mass derived from the X-ray surface brightness profile, provides a
high-quality proxy for the total mass.  The simulations, using different
codes, with or without including non-gravitational heating and cooling
of the cluster gas, and with different numerical techniques for treating
these effects, all show that $M_{\rm tot} \propto Y_X^\alpha$, with
$\lesssim 10\%$ scatter and a slope very close to the prediction of
self-similar theory, $\alpha=3/5$ (see \cite{2006ApJ...650..128K} and
later works). The low scatter in the $M_{\rm tot}-Y_X$ relation is
therefore a very \emph{robust} theoretical prediction, and the only
prediction we need to implement the test outlined below. The minimal
scatter in the $M_{\rm tot}-Y_X$ (and also $M_{\rm tot} - M_{\rm gas}$)
relations is confirmed by Chandra observations (e.g.,
\cite{2008MNRAS.383..879A,2008arXiv0805.2207V}).

Weak lensing techniques have a lower limit on the accuracy of mass
measurements for individual clusters of $20-30\%$, due to projection
effects. This scatter is too large for ``precision cosmology'' with the
cluster mass function. However, {\it on average}, weak lensing masses
are free of bias \cite{2009arXiv0901.3434C}. By combining the X-ray and
lensing approaches, and drawing on their individual strengths, one can
obtain mass measurements for samples of clusters that are both low in
intrinsic scatter and are bias free.

Once the systematic scatter in the X-ray mass proxy has been reduced to
$\lesssim10\%$, it will have only a small effect on cosmological
measurements from the cluster mass function.  The dominant uncertainties
are then associated with the weak lensing data and establishing the
\emph{normalization} of the $M_{\rm tot}-Y_X$ relation in each redshift
bin.

Observational calibration of the $M_{\rm tot}-Y_X$ relation is
essential. These cannot (currently) be predicted by theory with percent
level accuracy. The necessary weak lensing measurements will come from
survey data collected by ground and space-based projects like
Pan-STARRS, DES, LSST, JDEM and JWST, but also from targeted ground- and
space-based observations. The capabilities of 6m-class telescopes such
as Magellan or Subaru are adequate for average weak lensing measurements
out to $z\sim1$; beyond that, some kind of space-based data will be
required.  It could be either JWST pointings or survey data from JDEM or
EUCLID.  Assuming that the weak lensing data will provide
$M_{\text{tot}}$ with 30\% scatter and minimal average bias, then by
observing $\sim 100$ clusters in each redshift bin we will normalize the
$Y_X-M$ relation at that redshift to $\sim 3\%$. Given 3\% accuracy in
the normalization of the $Y_X-M$ relation, one can derive the linear
perturbation amplitude at this redshift to $1\%$ accuracy. If one were
to conservatively assume a factor of two degradation in these
measurements (representing a `pessimistic' scenario), then the same data
will still constrain the linear perturbation amplitude to 2\% accuracy
at each redshift.

Both the weak lensing and X-ray components are essential to this work.
If one has only weak lensing masses for the clusters, one cannot
accurately reconstruct the mass functions because of the large and
unavoidable $\sim 30\%$ scatter in the individual mass measurements.
Conversely, if one has only precise, X-ray measured $Y_X$ or $M_{\rm
  gas}$ parameters, it would be hard to control systematic biases in the
mass at the level sufficient for the 1--2\% growth measurements. Only
the combination of the two techniques circumvents these problems.

\subsection{The cluster samples}

\noindent Future, scheduled X-ray and SZ surveys will easily provide the
samples of clusters required for this work.  For example, eROSITA will
carry out a sensitive all-sky X-ray survey that will detect $\sim
200,000$ clusters, and this sample will have $>100$ clusters per $\Delta
z=0.1$ bin out to $z=1.5$. The serendipitous cluster catalog constructed
by IXO will probe two orders of magnitude lower in X-ray flux over a
smaller area and, together with SZ experiments, extend the target list
to $z=2$ and beyond. Proposed next-generation X-ray survey missions like
WFXT would extend these surveys yet further.

Selecting 100 massive, X-ray bright clusters in each $\Delta z=0.1$
redshift bin spanning the range $0<z<2$ gives a sample of 2000 clusters
requiring weak lensing and X-ray followup. Precise spectroscopic
redshifts will automatically be provided for each cluster by the X-ray
observations, but can also be obtained in a dedicated optical followup
program. 

The effective area of the survey-optimized X-ray telescopes telescopes
will be insufficient to measure $Y_X$, even with deep pointed
observations, for clusters at $z\geq 0.8$.  At higher redshifts, the
required $Y_X$ measurements will only be possible with a powerful
observatory such as IXO. Detailed exposure time estimates show that
$\approx 10$~Msec of IXO observing time will be required to carry out
this program. This is moderate, but by no means prohibitive, investment
of observing time over the lifetime of the mission, and will provide a
cosmological measurement of fundamental importance. The expected,
reconstructed structure growth history is illustrated in Fig.~1. The
redshift range of $0<z<2$ spans the entire epoch of accelerated
expansion. At the highest $z$, these studies will dovetail into planned
Ly-$\alpha$ forest observations.

\begin{figure}[t]
  \centerline{\includegraphics[width=0.99\linewidth]{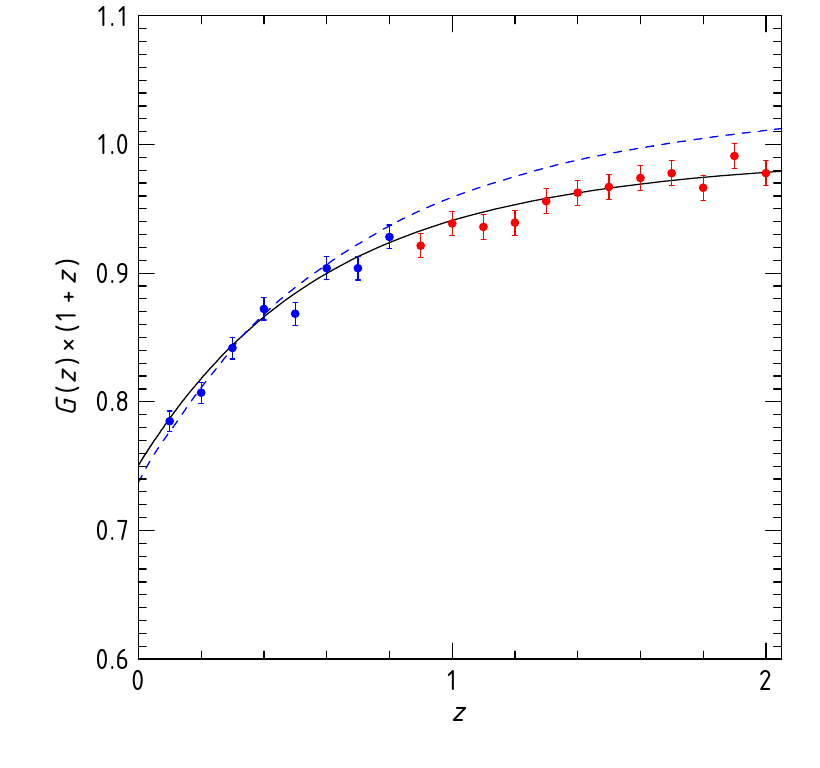}}
  \vspace*{-5mm}
  \caption{The normalized growth factor of density perturbations,
$G(z)$, constructed from follow-up X-ray and weak lensing observations
of 2000 clusters detected in sensitive X-ray surveys.  The extension
of $G(z)$ measurements to $z=2.0$ will be possible only with the high
sensitivity of IXO. The high-$z$ ($z>0.8$) data points are crucial for
testing non-GR models of cosmic acceleration. For example, the dashed
line shows the $G(z)$ function predicted for a DGP model with the same
expansion history as the quintessence model depicted by the solid
curve. }
\end{figure}

\subsection{Expected results from the $G(z)$ measurement}

Growth of structure, $G(z)$, data are highly complementary to
cosmological expansion history measurements in constraining, for
example, the dark energy equation of state. For illustration, we have
computed the combined constraints from a SNAP-like SNIa experiment
(adopted from \cite{2006astro.ph..9591A}) together with the $G(z)$ data
from the combined X-ray+weak lensing studies discussed above. The
results are shown in Fig.2. Because the distance- and growth-based
constraints are nearly orthogonal, their combination improves the
equation of state uncertainties by a factor of 2.5. Cluster growth of
structure data will provide a vital complement to the JDEM mission,
especially if that mission emphasizes $d(z$) measurements. 

Regardless of the accuracy in the equation of state measurements
achieved by JDEM, the next big question will be whether the cosmic
acceleration is caused by a physical scalar field or modifications of
General Relativity on large scales. The X-ray cluster data will be
crucial for testing such models because they, in general, significantly
modify the growth factor with respect to GR dark energy models with
similar distance-redshift relations. For example, the $d(z)$ relation
for a DGP modified gravity model (REF) can be almost indistinguishable
from the $d(z)$ of a quintessence model with $w\simeq-0.8$.  The growth
factor, however, is substantially different, as illustrated by the solid
and dashed lines in Fig.1. A DGP-type modification of the growth history
will be easily detectable with the proposed IXO measurements at $z>1$.

\begin{figure}[t]
  \centerline{\includegraphics[width=0.99\linewidth]{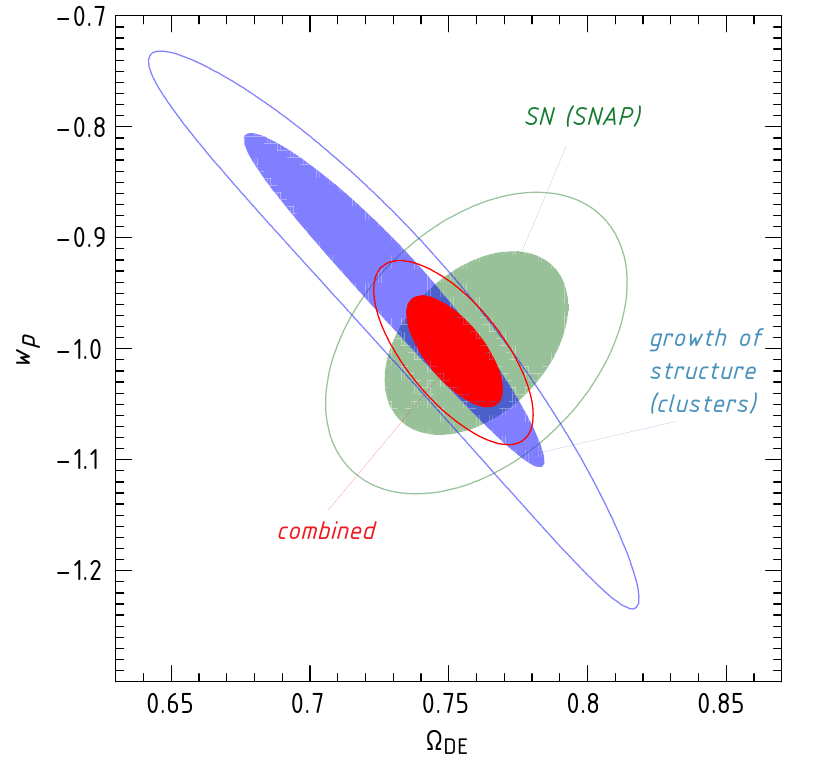}}
  \vspace*{-3mm}
  \caption{The improvement in the dark energy equation of state
    constraints obtained from the combination of distance-based
    techniques (projected results are shown for the SNAP SNIa
    experiment), and X-ray growth of structure measurements. Contours
    show the two-parameter 68\% and 95\% confidence regions, assuming
    Planck priors.  The combination of SNIa and X-ray growth of
    structure data improves the equation of state uncertainties by a
    factor of $\sim 2.5$.  $w_p$ is the value of equation of state at
    ``pivot'' redshift, where it is best constrained by the given
    experiment.}
\end{figure} 

A more quantitative demonstration of IXO capabilities in constraining
non-GR theories can be based on the so called ``growth index'', $\gamma$
\cite{2007PhRvD..75b3519H}.  For GR and a very wide range of
``physical'' dark energy models, such as quintessence, $\gamma$ is close
to 0.55.  Therefore, departures from General Relativity can be searched
for by detecting deviations in $\gamma$ from 0.55. For example,
$\gamma=0.68$ is predicted for the DGP model.  The projected $G(z)$
measurements in Fig.1 will constrain $\gamma$ to $\pm 0.022$ (0.045)
using cluster data only, and to $\pm 0.018$ (0.034) from combination of
the cluster and Planck data, assuming that the masses of light neutrinos
are known).  The values in parentheses are for the ``pessimistic''
calibration scenario in which projected uncertainties for mass
calibration are degraded by a factor of 2, as discussed above.  Huterer
\& Linder predict \cite{2007PhRvD..75b3519H} that $\gamma$ will be
constrained to $\pm0.044$ from combination of the supernovae and weak
lensing measurements from a SNAP-type mission combined with Planck CMB
priors.

\section{Probing the expansion history with $f_{\rm gas}(z)$}

The ratio of baryonic-to-total mass in clusters should closely match the
ratio of the cosmological parameters $\Omega_b/\Omega_m$ because the
matter content of the largest clusters of galaxies is expected to
provide a fair sample of the matter content of the Universe
\cite{1993Natur.366..429W}. The baryonic mass in clusters is dominated
by X-ray emitting gas, the mass of which exceeds the mass in stars by a
factor of $\sim 6$, with other sources of baryonic matter being
negligible. The combination of X-ray measurements of $f_{\rm gas}$ with
optical/near-IR estimates of the stellar mass, and determinations of
$\Omega_b$ and $H_0$ from e.g.\ CMB data, can therefore be used to
measure $\Omega_m$.

Measurements of $f_{\rm gas}$ as a function of redshift also probe the
acceleration of the Universe. This constraint originates both from the
fact that for the largest clusters $f_{\rm gas}$ is predicted to be a
near-invariant quantity with minimal intrinsic scatter
\cite{2007ApJ...655...98N,2008arXiv0808.1106F} and from the dependence
of the $f_{\rm gas}$ measurements (which are derived from the observed
X-ray temperature and density profiles, assuming hydrostatic
equilibrium) on the assumed distance to the clusters: $f_{\rm gas}
\propto d^{3/2}$. The latest results from this experiment
\cite{2008MNRAS.383..879A} using Chandra data for 42 hot, relaxed
clusters, give marginalized constraints of $\Omega_M=0.28\pm0.05$ and
$\Omega_\Lambda =0.86\pm0.22$ (Fig.3). The Chandra data confirm that the
Universe is accelerating at 99.99\% confidence, comparable in
significance to the best current SNIa data combined.  We emphasize that
systematic scatter remains $undetected$ in current Chandra $f_{\rm gas}$
data for hot, relaxed clusters, despite a weighted-mean statistical
distance error of only 5\% \cite{2008MNRAS.383..879A}. This compares
favorably with SNIa, where systematic scatter is detected at the 7\%
level in the individual distance estimates \cite{2007ApJ...659..122J}.

\begin{figure}
  \vspace*{-5mm}
  \centerline{\includegraphics[width=0.99\linewidth]{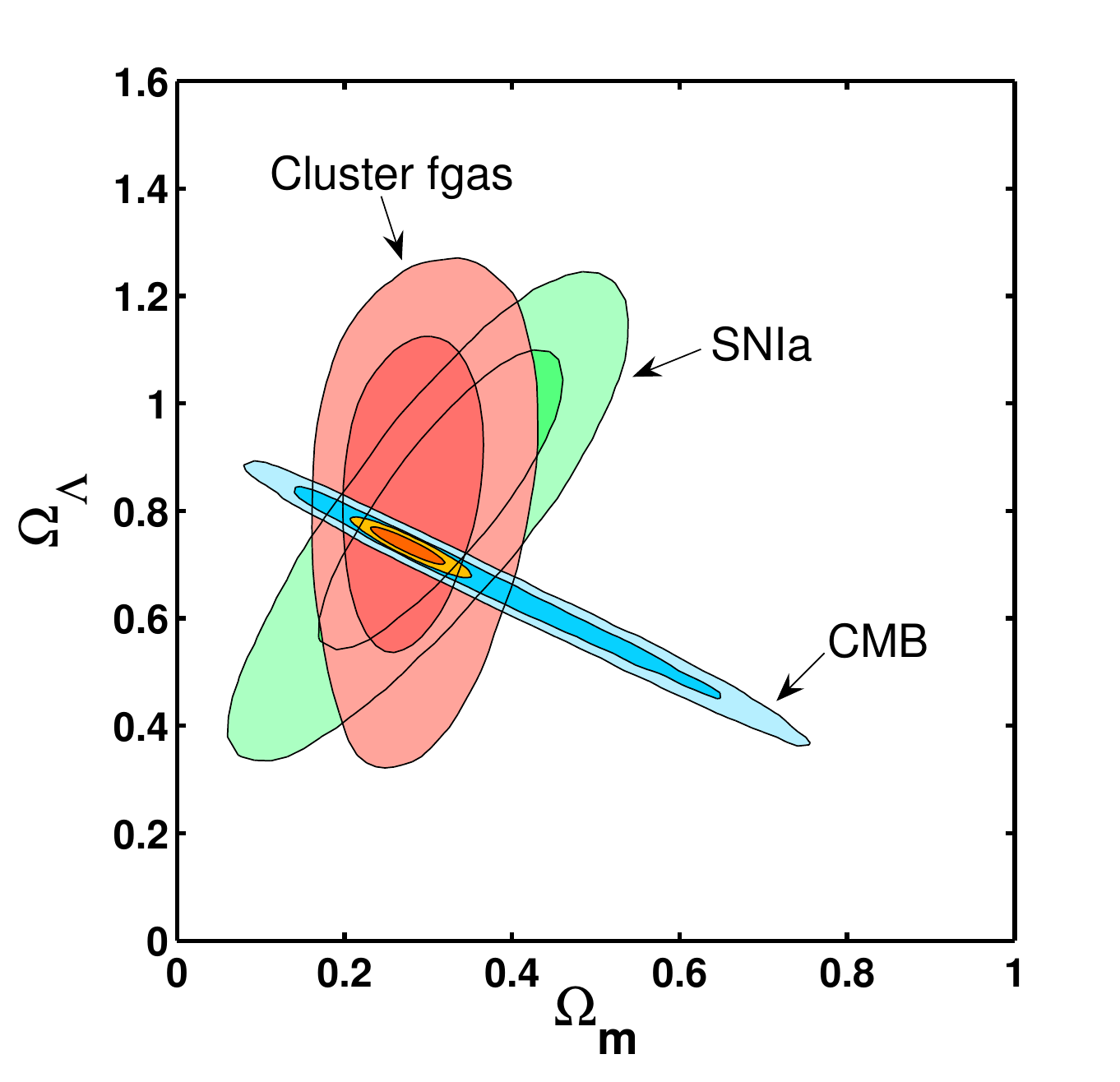}}
  \vspace*{-3mm}
  \caption{Constraints on $\Omega_M-\Omega_\Lambda$ from the current
    \emph{Chandra} $f_{\rm gas}$ measurements.}
\end{figure}

The prospects for $f_{\rm gas}$ studies with IXO have been studied in
detail in \cite{2008MNRAS.388.1265R}. An investment of $\sim10$\,Ms of
IXO time to measure $f_{\rm gas}$ to 5\% (corresponding to 3.3\%
accuracy in distance) in each of the 500 hottest, most X-ray luminous,
dynamically relaxed clusters detected in future cluster surveys, will be
sufficient to constrain cosmological parameters with a DETF
\cite{2006astro.ph..9591A} figure of merit (FoM) of 20--40. This range
in FoM spans pessimistic to optimistic assumptions regarding systematic
uncertainties \cite{2008MNRAS.388.1265R}. Similar cosmological
constraints would also be achievable by observing the best 250 clusters
for 40 ks each, on average; such a strategy may prove useful at higher
$z$ if the fraction of relaxed clusters is found to drop faster than
expected. Gravitational lensing data will again be used to pin-down the
mean hydrostatic mass bias in each redshift bin.  (These biases are not
expected to exceed 10\% for the largest, relaxed clusters.)

\begin{figure}
  \centerline{\includegraphics[width=0.85\linewidth]{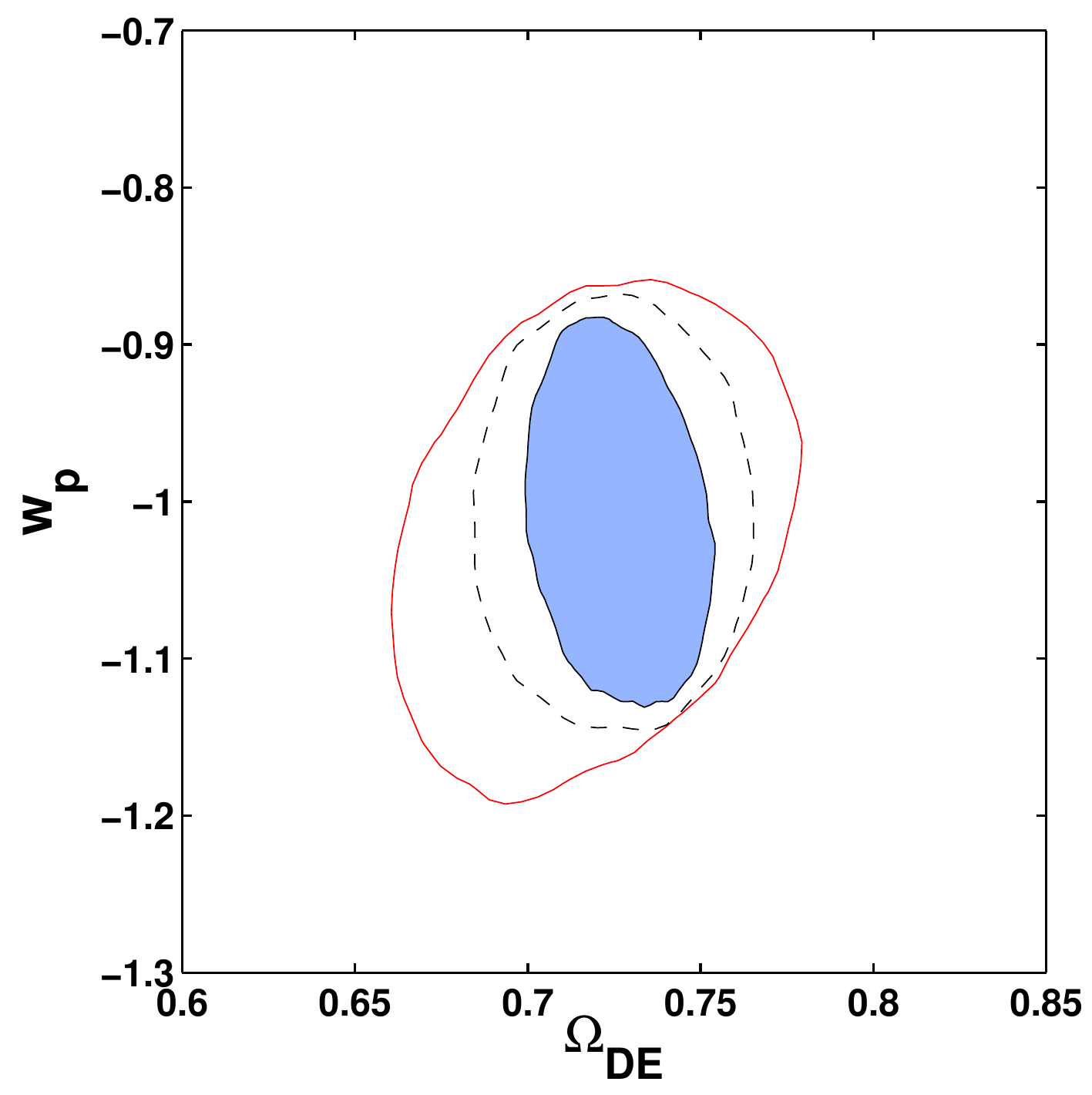}}
  \vspace*{-3mm}
  \caption{The projected 95\% confidence contours for the IXO $f_{\rm
      gas}$ experiment in the $\Omega_{\rm DE}-w_p$ plane for the
    default dark energy model and optimistic (2\%; blue, solid contour),
    standard (5\%; dashed contour) and pessimistic (10\%; red contour)
    allowances for systematics. The figure presented in an identical
    style to the DETF report\cite{2006astro.ph..9591A} to allow direct
    comparison with those results.}
\end{figure}

The constraints on the expansion history (see Fig.~4) from the IXO
$f_{\rm gas}$ experiment are comparable to, or exceed, those expected
for future `Stage IV' ground and space-based SNIa and BAO experiments
\cite{2006astro.ph..9591A}. In particular, the $f_{\rm gas}$ data are
expected to provide a very precise measurement of the mean matter
density, $\Omega_{\rm m}$.  Most importantly, the very different natures
of the astrophysics and systematics affecting the $f_{\rm gas}$, SNIa
and BAO experiments will ensure maximum robustness when the results are
combined.  The addition of follow-up SZ observations will boost the IXO
FoM still further, providing independent constraining power via the
classic `XSZ' experiment that combines X-ray and SZ measurements of the
Compton y-parameter \cite{2008MNRAS.388.1265R,2006ApJ...647...25B}.
Although less intrinsically powerful, than the $f_{\rm gas}$ test, the
XSZ experiment rests on different assumptions and has different
systematic uncertainties. The optimal IXO observing strategies for both
the $f_{\rm gas}$ and XSZ experiments are identical and will use the
same X-ray observations of the largest, dynamically relaxed
clusters\cite{2008MNRAS.388.1265R}.

\section{Summary}

\noindent A moderate investment of observing time with the International
X-ray Observatory to study high-redshift galaxy clusters detected in
future large-scale surveys, will provide cosmological measurements of
fundamental importance. IXO observations, combined with lensing
follow-up, will measure the perturbation growth factor from $z=0-2$ with
an accuracy comparable to, or possibly better than, that expected from
observations of cosmic shear with JDEM, and redshift-space distortions
with EUCLID.  The growth of structure data derived from clusters will
significantly improve our knowledge of the dark energy equation of state
and will aid in constraining non-GR models for cosmic acceleration. IXO
observations of the largest, dynamically relaxed clusters will provide a
powerful, independent measurement of the cosmological expansion history
using the apparent $f_{\rm gas}(z)$ trend. Systematic and statistical
errors from this technique are competitive with SNIa and BAO studies,
making the test extremely useful for improving the accuracy and
reliability of the geometric cosmological measurements planned for LSST
and JDEM.  Only by employing a range of powerful, independent
approaches, including those discussed here, can robust answers to
puzzles as profound as the origin of cosmic acceleration be expected.

\bibliographystyle{prop}
\bibliography{ixo}

\end{document}